\documentclass[aps, 10pt, pr,notitlepage, twocolumn, superscriptaddress,nofootinbib]{revtex4-2}

\usepackage{amsmath}
\usepackage{amssymb}
\usepackage{amsfonts}
\usepackage[utf8]{inputenc}
\usepackage[T1]{fontenc}
\usepackage{mathrsfs}
\usepackage{anyfontsize}
\usepackage{hyperref}
\usepackage[usenames,dvipsnames]{xcolor}

\usepackage{newtxtext}
\usepackage{newtxmath}
\usepackage{tensor}
\usepackage{bm}

\usepackage{graphicx}
\usepackage{epsfig}
\usepackage{epstopdf}

\usepackage{natbib}

\usepackage[normalem]{ulem}

\newcommand{\dF}{\frac{d\ln A}{d\chi}}

\def\jnl@style{\it}
\def\aaref@jnl#1{{\jnl@style#1}}

\def\aaref@jnl#1{{\jnl@style#1}}

\def\aj{\aaref@jnl{AJ}}                   
\def\apj{\aaref@jnl{ApJ}}                 
\def\apjl{\aaref@jnl{ApJ}}                
\def\apjs{\aaref@jnl{ApJS}}               
\def\apss{\aaref@jnl{Ap\&SS}}             
\def\aap{\aaref@jnl{A\&A}}                
\def\aapr{\aaref@jnl{A\&A~Rev.}}          
\def\aaps{\aaref@jnl{A\&AS}}              
\def\mnras{\aaref@jnl{Mon.~Not.~Roy.~Astron.~Soc.}}             
\def\prd{\aaref@jnl{Phys.~Rev.~D}}        
\def\prc{\aaref@jnl{Phys.~Rev.~C}}  
\def\prl{\aaref@jnl{Phys.~Rev.~Lett.}}    
\def\qjras{\aaref@jnl{QJRAS}}             
\def\skytel{\aaref@jnl{S\&T}}             
\def\ssr{\aaref@jnl{Space~Sci.~Rev.}}     
\def\zap{\aaref@jnl{ZAp}}                 
\def\nat{\aaref@jnl{Nature}}              
\def\aplett{\aaref@jnl{Astrophys.~Lett.}} 
\def\apspr{\aaref@jnl{Astrophys.~Space~Phys.~Res.}} 
\def\physrep{\aaref@jnl{Phys.~Rep.}}      
\def\physscr{\aaref@jnl{Phys.~Scr}}       
\def\commat{\aaref@jnl{Comm.~Math.~Phys.}}              
\def\science{\aaref@jnl{Science}}               
\def\cqg{\aaref@jnl{Classical Quant.~Grav.}}            
\def\jpcs{\aaref@jnl{JPCS}}                                     
\def\ijmpd{\aaref@jnl{Int.~J.~Mod.~Phys.~D}}                    
\def\grg{\aaref@jnl{Gen.~Relat.~Gravit.}}               
\def\rpp{\aaref@jnl{Rep.~Prog.~Phys.}}          
\def\npa{\aaref@jnl{Nucl.~Phys.~A}}        
\def\lrr{\aaref@jnl{Living Rev.~Rel.}}                   
\def\jcap{\aaref@jnl{J.~Cosmology Astropart.~Phys.}}    
\def\rmp{\aaref@jnl{Rev.~Mod.~Phys.}}   

\allowdisplaybreaks[1]

\begin{document}
	
	\title{Nonlinear evolution and non-uniqueness of scalarized neutron stars}
	
	
	\author{Hao-Jui Kuan}
	\email{hao-jui.kuan@uni-tuebingen.de}
	\affiliation{Theoretical Astrophysics, Eberhard Karls University of T\"ubingen, T\"ubingen 72076, Germany}
	\affiliation{Department of Physics, National Tsing Hua University, Hsinchu 300, Taiwan}
	
	\author{Jasbir Singh}
	\email{jasbir.singh@inaf.it}
	\affiliation{Theoretical Astrophysics, Eberhard Karls University of T\"ubingen, T\"ubingen 72076, Germany}
	\affiliation{Department of Physics, University Of Trieste, via Tiepolo 11, 34143 Trieste, Italy}
	\affiliation{Department of Space, Earth and Environment, Chalmers University of Technology, 41293 Gothenburg, Sweden}
	\affiliation{INAF- Astronomical Observatory of Trieste, via Tiepolo 11, 34143 Trieste, Italy}

	\author{Daniela D. Doneva}
	\email{daniela.doneva@uni-tuebingen.de}
	\affiliation{Theoretical Astrophysics, Eberhard Karls University of T\"ubingen, T\"ubingen 72076, Germany}
    \affiliation{INRNE - Bulgarian Academy of Sciences, 1784  Sofia, Bulgaria}

	\author{Stoytcho S. Yazadjiev}
	\email{yazad@phys.uni-sofia.bg}
	\affiliation{Theoretical Astrophysics, Eberhard Karls University of T\"ubingen, T\"ubingen 72076, Germany}
	\affiliation{Department of Theoretical Physics, Faculty of Physics, Sofia University, Sofia 1164, Bulgaria}
	\affiliation{Institute of Mathematics and Informatics, 	Bulgarian Academy of Sciences, 	Acad. G. Bonchev St. 8, Sofia 1113, Bulgaria}
	
	\author{Kostas D. Kokkotas}
	\email{kostas.kokkotas@uni-tuebingen.de}
	\affiliation{Theoretical Astrophysics, Eberhard Karls University of T\"ubingen, T\"ubingen 72076, Germany}

	\begin{abstract}
		It was recently shown, that in a class of tensor-multi-scalar theories of gravity with a nontrivial target space metric, there exist scalarized neutron star solutions. An important property of these compact objects is that the scalar charge is zero and therefore, the binary pulsar experiments can not impose constraints based on the absence of scalar dipole radiation. Moreover, the structure of the solutions is very complicated. For a fixed central energy density up to three neutron star solutions can exist -- one general relativistic and two scalarized, that is quite different from the scalarization in other alternative theories of gravity. In the present paper we address the stability of these solutions using two independent approaches -- solving the  linearized radial perturbation equations and performing nonlinear simulations in spherical symmetry. The results show that the change of stability occurs at the maximum mass models and all solutions before that point are stable. This leads to the interesting consequence that there exists a stable part of the scalarized branch close to the bifurcation point where the mass of the star increases with the decrease of the central energy density.  
	\end{abstract}
	
	
	\maketitle
	
	\section{Introduction}
	Perhaps the most widely studied models of compact stars in alternative theories of gravity are the scalarized neutron stars in the Damour-Esposito-Farese (DEF) scalar-tensor theory of gravity \cite{Damour1993,Damour:1996ke}. The reason is that these were the first models that offered the possibility to have a theory perturbatively equivalent to General Relativity (GR), and thus no constraints from the weak field observations can be imposed, while still allowing for large deviations in the strong gravitation regime due to a nonlinear development of the scalar field. This mechanism for development of a nontrivial scalar field is possible for other compact objects, such as black holes \cite{Stefanov2008, Doneva2010, Cardoso2013,Doneva_2018,Silva_2018,Antoniou:2017acq}, but it requires either some not very realistic astrophysical conditions, or further modifications of the Hilbert-Einstein action such as the inclusion of curvature invariants. For neutron stars, the matter itself can act as a source of the scalar field due to the nonzero trace of the energy momentum tensor, and thus scalarized neutron stars became naturally the primary target for investigating the possible effects of nontrivial scalar hair and its observational implications.
	
	Scalarization indeed can produce very large deviations from GR, but in the standard DEF model it leads to the emission of scalar dipole radiation that is severely limited by the binary pulsar observations \cite{Damour:1996ke,Lazaridis:2009kq,Antoniadis:2013pzd,Freire:2012mg,Shao:2017gwu}. An elegant way to evade these constraints is to consider a nonzero scalar field mass, that suppresses the scalar dipole radiation \cite{Popchev2015, Ramazanoglu:2016kul, Yazadjiev:2016pcb,Rosca-Mead:2020bzt}. Another more sophisticated and also viable approach is to allow for the presence of multiple scalar fields. This is possible in the tensor-multi-scalar theories (TMST) of gravity that are the generalization of the standard scalar-tensor theories to multiple scalar field. These theories are mathematically self-consistent and well posed,  and can pass through all known experimental and observational tests \cite{Damour_1992,Horbatsch_2015,Yazadjiev:2019oul,Doneva:2019krb,Doneva:2019ltb}. Moreover, (quantum motivated) higher-order generalizations of GR often predict the existence of multiple scalar fields \cite{Damour_1992, Gottlober_1990}.  
	
	In TMST different kinds of interesting compact objects can be constructed including solitons \cite{Yazadjiev:2019oul,Collodel:2019uns}, mixed soliton-fermion stars \cite{Doneva:2019krb}, topological and scalarized neutron stars \cite{Doneva:2019ltb, Doneva:2020stability_tns, Doneva:2020ntsns}. The variety of solutions is controlled mainly by the choice of target space for the scalar fields $\varphi$ and the metric defined on it, and the choice of the map $\varphi : \text{\it spacetime} \to \text{\it target space}$. In particular, for a nontrivial map $\varphi : \text{\it spacetime} \to \text{\it target space}$ where the \emph{target space} is a maximally symmetric 3-dimensional space ($\mathbb{S}^3$, $\mathbb{H}^3$ or $\mathbb{R}^3$), there exists non-topological, spontaneously scalarized neutron stars in this theory \cite{Doneva:2020ntsns}. These are mathematically similar to topological neutron stars \cite{Doneva:2019ltb}, but with an important difference: the value of the scalar field at the center of the star is zero and thus the topological charge vanishes. A very important property of these solutions is that they have a zero scalar charge and thus no emission of scalar dipole radiation is possible. Therefore, the strong observational constraints on the standard scalar-tensor theories obtained on the basis of the binary pulsar observations simply do not apply for the TMST under consideration that allows for strong possibly observable deviations from GR.
	
	As discovered in \cite{Doneva:2020ntsns}, the scalarized TMST neutron stars show a very interesting property related to the uniqueness of the solutions. This constitutes in the fact that for a fixed central energy density up to three neutron star solutions can exist -- one GR solution with zero scalar field  and up to two scalarized solutions. This is in sharp contradiction with the standard scalar-tensor theories \cite{Damour1993} where only one scalarized neutron star solution can exist for a given central energy density. The preliminary stability analysis performed in \cite{Doneva:2020ntsns} based on the turning point method, suggested that all three of the solutions are stable (where exist). In the present paper we go further by performing a stability analysis (both a linear and nonlinear one) in order to determine the (in)stability of the scalarized neutron stars. Radial perturbations of neutron stars in scalar-tensor theories have already been studied in \cite{Sotani:2014,Mendes:2018} while the linear stability of TMST for topological neutron stars was examined in \cite{Doneva:2020stability_tns}. 
	
	
	In section \ref{section:theory}, we give a brief overview of the theory of scalarized neutron stars and in section \ref{section:bkg_sol} we present the background neutron star solutions. The stability of these solutions is examined in sections \ref{section:linear} and \ref{section:nonlinear} in the linear and nonlinear regimes respectively. Finally, the conclusions are presented in section \ref{section:concl}.

	\section{Neutron stars in tensor-multi-scalar theories of gravity} \label{section:theory}
	
	In this section, we will briefly describe the basics of TMST and especially the subclass of these theories that allows for the construction of scalarized neutron stars. For a more extensive discussion, we refer the reader to the original paper where these solutions where constructed  \cite{Doneva:2019ltb}.
	
	The most general action of TMST in the Einstein frame can be written in the form \cite{Damour_1992,Horbatsch_2015}: 
	\begin{eqnarray}\label{Action}
		S=&& \frac{1}{16\pi G_{*}}\!\int\!\! d^4\sqrt{-g}\left[R - 2g^{\mu\nu}\gamma_{ab}(\varphi)\nabla_{\mu}\varphi^{a}\nabla_{\nu}\varphi^{b} - 4V(\varphi)\right]  \nonumber \\
		&&+ S_{m}(A^{2}(\varphi) g_{\mu\nu}, \Psi_{m}),
	\end{eqnarray}
	where $G_{*}$ is the bare gravitational constant, $\nabla_{\mu}$ and $R$ are the covariant derivative and Ricci scalar respectively, both associated with $g_{\mu\nu}$. $V(\varphi)\ge 0$ denotes the potential of the scalar fields and $\Psi_{m}$ represents collectively the matter fields. The theory is equipped with $N$ scalar fields $\varphi_a$ that define a map $\varphi : \text{\it spacetime} \to \text{\it target space}$, where the $\text{\it target space}$ is a $N$-dimensional Riemannian manifold ${\cal E}_{N}$ with $\gamma_{ab}(\varphi)$ as a positively definite metric defined on it. The function $A(\varphi)$ is the conformal factor connecting the metrics in the Einstein frame ($g_{\mu\nu}$) and the physical Jordan frame ($\tilde{g}_{\mu\nu}$) via the relation ${\tilde g}_{\mu\nu}= A^2(\varphi) g_{\mu\nu}$. In our calculations we will adopt the Einstein frame for mathematical simplicity while all final quantities will be transformed to the  physical frame. Unless otherwise specifies, tilde will denote the quantities in the Jordan frame.
	
	By varying the action (\ref{Action}) with respect to the metric and the scalar fields, we obtained the following field equations in the Einstein frame:
	\begin{align}\label{eq:fieldeq}
		&R_{\mu\nu}= 2\gamma_{ab}(\varphi) \nabla_{\mu}\varphi^a\nabla_{\nu}\varphi^b + 2V(\varphi)g_{\mu\nu} \nonumber \\
		&\hskip 1.0cm + 8\pi G_{*} \left(T_{\mu\nu} - \frac{1}{2}T g_{\mu\nu}\right), \nonumber \\
		&\nabla_{\mu}\nabla^{\mu}\varphi^a = - \gamma^{a}_{\, bc}(\varphi)g^{\mu\nu}\nabla_{\mu}\varphi^b\nabla_{\nu}\varphi^c 
		+ \gamma^{ab}(\varphi) \frac{\partial V(\varphi)}{\partial\varphi^{b}} \nonumber \\
		&\hskip 1.6cm -  4\pi G_{*}\gamma^{ab}(\varphi)\frac{\partial\ln A(\varphi)}{\partial\varphi^{b}}T,
	\end{align}
	where $\gamma^{a}_{\, bc}(\varphi)$ denotes the Christoffel symbols of the target space metric $\gamma_{ab}(\varphi)$. The Einstein frame energy-momentum tensor $T_{\mu\nu}$ satisfies the following conservation relation:
	\begin{eqnarray}\label{eq:eulereinstein}
		\nabla_{\mu}T^{\mu}_{\nu}= \frac{\partial \ln A(\varphi)}{\partial \varphi^{a}}T\nabla_{\nu}\varphi^a .
	\end{eqnarray}
	The energy-momentum tensor in the Jordan frame is given by $\tilde{T}_{\mu\nu}=A^{-2}(\varphi)T_{\mu\nu}$. We only consider perfect fluid stars in our analysis and thus the energy density, the
	pressure and the 4-velocity are connected in the two frames by $\varepsilon=A^{4}(\varphi){\tilde \varepsilon}$, $p=A^{4}(\varphi){\tilde p}$ and $u_{\mu}=A^{-1}(\varphi) {\tilde u}_{\mu}$ respectively.

	Since we are interested in static, spherically symmetric and asymptotically flat solutions, the metric takes the following general form
	\begin{eqnarray}
		ds^2= - e^{2\Gamma(r)}dt^2 + e^{2\Lambda(r)}dr^2 + r^2(d\theta^2  + \sin^2\theta d\phi^2).
		\label{eq:metric}
	\end{eqnarray} 
	where metric function $\Lambda(r)$ is related to the mass enclosed within the circumferential radius $r$ via
	\begin{align}
		e^{-2 \Lambda} = 1 - \frac {2 m(r)} {r}.
	\end{align}
	The 4-velocity of a generic fluid moving radially is
	\begin{align}
		\tilde{u}^{\mu} = \frac{1}{\sqrt{1-v^2}}(e^{-\Gamma}\partial_{t} + ve^{-\Lambda}\partial_{r}),
	\end{align}
	with the characteristic strength $v$.

	The simplest setup that can lead to the existence of the desired scalarized solutions is the following \cite{Doneva:2020ntsns}. We consider three scalar fields $\varphi_a=\{\chi,\Theta,\Phi\}$, with the target space manifold being $\mathbb{S}^3$, $\mathbb{H}^3$ or $\mathbb{R}^3$. Thus the 3-dimensional target space metric takes the following form:
	\begin{eqnarray}
		\gamma_{ab}d\varphi^a d\varphi^b= a^2\left[d\chi^2 + H^2(\chi)(d\varTheta^2 + \sin^2\varTheta d\Phi^2) \right],
	\end{eqnarray}
	where $\varTheta$ and $\Phi$ are the standard angular coordinates on the 2-dimensional sphere $\mathbb{S}^2$ and the parameter $a$ is related to the curvature of $\mathbb{S}^3$ and $\mathbb{H}^3$. The function $H(\chi)$ represents the target space geometry: for spherical geometry $\mathbb{S}^3$, $H(\chi)=\sin\chi$; for hyperbolic geometry $\mathbb{H}^3$, $H(\chi)=\sinh\chi$; and finally for flat geometry $\mathbb{R}^3$, $H(\chi)=\chi$. We will only consider theories where the coupling function $A(\varphi)$ and the potential $V(\varphi)$ depend only on $\chi$, which in turn allows the equations for $\Theta$ and $\Phi$ to separate. This guarantees that the spacetime will be spherically symmetric in both the Einstein and the Jordan frames for the ansatz defined below.
	
	In this paper we choose a nontrivial map $\varphi$ such that the field $\chi$ is assumed to depend on the radial coordinate $r$ while $\varTheta$ and $\Phi$ are independent from $r$ and are given by $\Theta=\theta$ and $\Phi=\phi$ \cite{Doneva:2019ltb,Doneva:2020ntsns}. This ansatz is compatible with the spherical symmetry and in addition, ensures that the equations for $\varTheta$ and $\Phi$ are satisfied. 
	
	Using the ansatz stated above and the general form of the field equations \eqref{eq:fieldeq}, the dimensionally reduced field equations governing the neutron star equilibrium solutions can be derived. Since they are somewhat lengthy and also not the main focus of the present paper, we will not present them here and refer the reader to \cite{Doneva:2020ntsns}. They have to be supplemented with boundary conditions and we consider the standard ones -- regularity at the center of the star and asymptotic flatness.Thus we impose $\Gamma(\infty)=0$, $\Lambda(\infty)=0$ and $\chi(\infty)=0$, while at the stellar center $\Lambda(0)=0$ and $\chi(0)=0$. As a matter of fact for a target space being $\mathbb{S}^3$, the scalar field $\chi$ can have a more general boundary condition at the center  $\chi(0)=n\pi$ with $n\in \mathbb{Z}$ being the stellar topological charge \cite{Doneva:2019ltb,Doneva:2020stability_tns}. In the present paper, though, we will be focusing only on non-topological scalarized neutron stars and thus consider $n=0$. 
	
	At infinity  the scalar field $\chi$ behaves as    
	\begin{eqnarray}
		\chi\approx \frac{{\rm const}}{r^2} + O(1/r^3).
	\end{eqnarray}
	In this expansion, the $1/r$ term is missing and thus the scalar charge is zero. This implies that these starts do not emit any scalar dipole radiation and therefore they comply with the binary pulsar observations by construction. Furthermore, since the leading order term in the expansion is proportional to $1/r^2$, the ADM masses in both frames are the same.
	
	\section{The background solutions}
	\label{section:bkg_sol}
	Here, we will briefly present the behavior of the background solutions that will be later evolved. More details can be found in \cite{Doneva:2020ntsns}. 
	
	Since we want to construct scalarized neutron stars, the conformal factor function $A(\chi)$ has to be chosen accordingly. More precisely, it should satisfy the following conditions
	\begin{eqnarray}\label{CA}
		\frac{\partial A}{\partial \chi}(0)=0, \,\,\,  \frac{\partial^2 A}{\partial \chi^2}(0)\ne 0.
	\end{eqnarray}
	Taking these conditions into account, we employ the following standard form of the conformal factor 
	\begin{equation}
		A(\chi)=e^{\beta \alpha(\chi)},
	\end{equation}
	where $\alpha(\chi)$ is a function of the scalar field and can be, for example, a periodic function such as $\sin^2{\chi}$, or simply $\chi^2$. It can be easily shown that the coupling function with these choices for $\alpha(\chi)$ satisfies the conditions \eqref{CA}. 
	
	The dimensionally reduced field equations together with the above mentioned boundary conditions are solved numerically using a shooting method. The shooting parameters are the central values of the scalar field derivative $(d\chi/dr)(0)$ and the metric function $\Gamma(0)$. They are determined by the conditions that $\chi$ and $\Gamma$ tend to zero at (numerical) infinity.
	
\begin{figure}
		\hspace*{-0.4cm}  \includegraphics[scale=0.5]{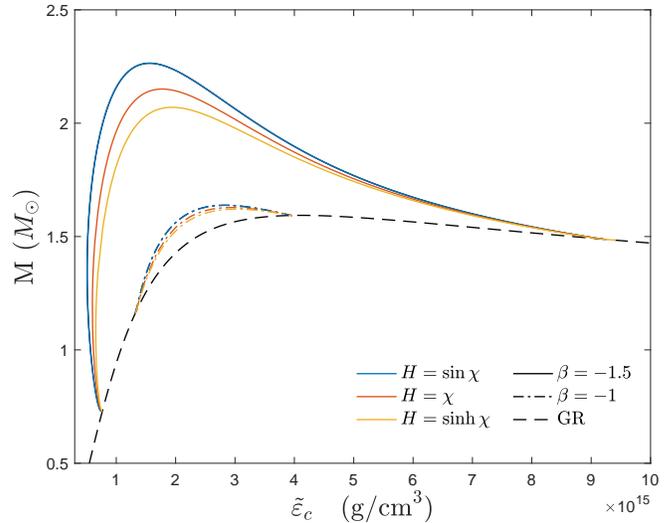}
		\caption{The mass as a function of the central energy density for the fundamental branch of scalarized neutron stars possessing nodeless scalar field. Solutions for the cases with $A(\chi)=\exp(\beta\sin^{2}\chi)$ and $H(\chi)=\{\sin\chi,\chi,\sinh(\chi)\}$ are shown. The values of the parameters are fixed to $a^2=0.1$ and $\beta=\{-1,-1.5\}$. The neutron stars with zero scalar field are plotted with a black line.}
		\label{fig:M_rhoc}
	\end{figure}

	Fig. \ref{fig:M_rhoc} shows the neutron star mass $M$ as a function of the central energy density $\tilde{\varepsilon}_c$  for a conformal factor  $A(\chi)=\exp(\beta\sin^{2}\chi)$ and the three possible choices of $H(\varphi)$. In this figure, we used a hybrid  equation of state (EOS) to account for the stiffening of the matter at nuclear density $\tilde{\rho}_{\rm nucl}=2\times 10^{14} g\, cm{^{-3}}$, where the pressure and the internal energy are given by 
	\begin{eqnarray}
		&&\tilde{p}= K_1 \tilde{\rho}^{\Gamma_1}, \;\;\;  \tilde{\varepsilon}_i= \frac{K_1}{\Gamma_1 -1 } \tilde{\rho}^{\Gamma_1 - 1}, \;\; {\rm for} \;\; \tilde{\rho} \le \tilde{\rho}_{\rm nucl} , \label{eq:EOS1}\\
		&&\tilde{p}= K_2 \tilde{\rho}^{\Gamma_2}, \;\;\;  \tilde{\varepsilon}_i= \frac{K_2}{\Gamma_2 -1 } \tilde{\rho}^{\Gamma_2 - 1}, \;\; {\rm for} \;\; \tilde{\rho} > \tilde{\rho}_{\rm nucl} , \label{eq:EOS2}.
	\end{eqnarray}
	The energy density and the internal energy are related to each other via $\tilde{\varepsilon}=\tilde{\rho}(1+\tilde{\varepsilon}_{i})$.
	This equation of state clearly does not reach the two solar mass barrier, but it was widely used for example in the nonlinear simulations of stellar evolution in scalar-tensor theories \cite{Janka93,Zwerger97,Dimmelmeier02,OConnor:2009iuz,Gerosa:2016fri}. Since our nonlinear code for examining the stability is based on \cite{OConnor:2009iuz,Gerosa:2016fri} we decided to keep this EOS for consistency. We have performed calculations for other piecewise polytropic EOS \cite{PPA_Read} and the results remain qualitatively the same.
	
	As one can easily check, the GR neutron star solutions with zero scalar field are always solutions of the field equations \eqref{eq:fieldeq} if it obeys the conditions  \eqref{CA}. At a certain energy density $\tilde{\varepsilon}_c^{\rm crit}$ a nontrivial scalar field develops and the solutions bifurcate from GR. As discussed in \cite{Doneva:2020ntsns}, $\tilde{\varepsilon}_c^{\rm crit}$ is controlled only by the values of $\beta/a^2$ and it is independent on the particular form of the coupling function (as long as it allows for scalarization of course). These scalarized solutions coexist with the GR solutions indicating non-uniqueness and they are energetically favorable. However, at a particular higher value of the central energy density, the scalarized branch of solutions merges again with the GR one and the neutron stars with nonzero scalar field cease to exist. With the decrease of $\beta/a^2$ the range of central energy densities, where scalarized solutions exist, gets larger  and the deviations from GR increase. It is interesting to note a well known fact in scalar-tensor theories -- the scalarization increases the maximum mass and thus an EOS that in GR leads to neutron star masses lower than the two solar mass barrier, can reach above this threshold in the presence of nontrivial scalar field. What is different from all the other examples of scalarized neutron stars in standard (massless) scalar-tensor theories, though, is that for the TMST solutions the scalar charge is zero. Thus, they can not be constrained by the binary pulsar observations and allow for large deviations from GR.  
	
	For larger values of $\beta$ (e.g. $\beta=-1$), the mass of the scalarized neutron stars increases monotonically as the central energy density increases till the maximum of mass is reached and after that the mass keeps on decreasing until the branch merges with the GR solutions. On the other hand, for lower values of $\beta$, after the first bifurcation point the mass of the scalarized neutron stars increases whereas $\tilde{\varepsilon}_c$ decreases. This happens until a minimum value of $\tilde{\varepsilon}_c$ is reached and after that the behavior of the branch is similar to the larger $\beta$ case. This different behavior of the smaller $\beta$ branch implies that at certain lower values of $\tilde{\varepsilon}_c$, there exist simultaneously three solutions -- two scalarized ones and one solution with zero scalar field, which indicates non-uniqueness. This is a new results that has not been observed in standard scalar-tensor theories.
	
	We should note that the particular choice of the coupling function only deforms the scalarized branch, while keeping the position of the bifurcation points unaltered \cite{Doneva:2020ntsns}. That is why, even though we have presented here the $M(\tilde{\varepsilon}_c)$ dependence only for $A(\chi)=\exp(\beta\sin^{2}\chi)$, the results are qualitatively the same for other couplings such as $A(\chi)=\exp(\beta \chi^2)$.
	
	Below we will study the stability of the scalarized solutions with two independent approaches -- by examining the linearized field equations and by considering the full system of nonlinear field equations in spherical symmetry. Even though the former approach should in principle constitute a subclass of the latter one, we have decided to apply both of them in order to have an independent verification of (in)stability especially taking into account the observed very interesting non-uniqueness of solutions.
	
	\section{Linear Scheme}
	\label{section:linear}
	
	\subsection{Perturbation Equations}
	
	To derive the perturbation equations for the radial stability analysis, in the field equations we impose perturbations of the form
	\begin{equation}
		f(t,r)=f_0(r)+\delta f(t,r),
	\end{equation}
	where $f$ represents a perturbed variable which in our case is the metric functions, the Jordan frame pressure $\tilde{p}$ and energy density $\tilde{\varepsilon}$, and the scalar field $\chi$. The static background functions are denoted by a subscript "$0$" in $f_0$ and the time dependent radial perturbations are represented by $\delta f$. As a matter of fact, the fluid perturbations can be expressed in terms of the Lagrangian displacement $\zeta=\zeta(t,r)$ as we will see below. 
	
	In a perturbed state, the star pulsates around the spherically symmetric equilibrium configuration, with the line element as
	\begin{equation}\label{pert_metric}
		ds^2 = -e^{2\Gamma_0 + 2\delta\Gamma}dt^2 + e^{2\Lambda_0 + 2\delta\Lambda}dr^2 + r^2(d\theta^2+\textup{sin}^2\theta d\varphi^2).
	\end{equation}
	The equations governing the fluid perturbation $\zeta$ and the scalar field perturbation $\delta\chi$ are given as
	\begin{widetext}
		\begin{align}
			&(\tilde{\varepsilon}_0+\tilde{p}_0)e^{2\Lambda_0-2\Gamma_0}\ddot{\zeta}+ (\tilde{\varepsilon}_0+\tilde{p}_0)\delta\Gamma'+[\Gamma_0'+\alpha(\chi_0)\chi_0'](\delta\tilde{\varepsilon}+\delta\tilde{p}) + \delta\tilde{p}' + \alpha(\chi_0)(\tilde{\varepsilon}_0+\tilde{p}_0)\delta\chi' + \tilde{\beta}(\chi_0)(\tilde{\varepsilon}_0+\tilde{p}_0)\chi_0'\delta\chi=0, \label{eq:WE1}\\
			&-e^{-2\Gamma_0}\ddot{\delta\chi}+e^{-2\Lambda_0}\delta\chi''+e^{-2\Lambda_0}\left[\Gamma_0'-\Lambda_0'+\frac{2}{r}\right]\delta\chi' + e^{-2\Lambda_0}\chi_0'[\delta\Gamma'-\delta\Lambda'] + \left[-\frac{2}{r^2}\left(\frac{\mathrm{d}}{\mathrm{d}\chi}H^2(\chi)\right)_{\chi_0}+\frac{2}{a^2}\partial_\chi V(\chi_0)\right. \notag \\
			&\hskip 1cm \left.- 8\pi G_*\frac{\alpha(\chi_0)}{a^2}A^4(\chi_0)(\tilde{\varepsilon}_0-3\tilde{p}_0)\right]\delta\Lambda -\left[\frac{1}{r^2}\left(\frac{\mathrm{d}^2}{\mathrm{d}\chi^2}H^2(\chi)\right)_{\chi_0}+\frac{1}{a^2}\partial^2_\chi V(\chi_0) + 4\pi G_*\frac{\beta(\chi_0)}{a^2}A^4(\chi_0)(\tilde{\varepsilon}_0-3\tilde{p}_0)\right. \notag \\ 
			&\hskip 1cm \left.+ 16\pi G_*\frac{\alpha^2(\chi_0)}{a^2}A^4(\chi_0)(\tilde{\varepsilon}_0-3\tilde{p}_0)\right]\delta\chi - 4\pi G_*\frac{\alpha(\chi_0)}{a^2}A^4(\chi_0)(\delta\tilde{\varepsilon}-3\delta\tilde{p})=0, \label{eq:WE2}
		\end{align}
	\end{widetext}
	where dot and prime represent derivatives with respect to time and radial coordinates, respectively, and $\alpha(\chi) = \frac{d\:\textup{ln}A(\chi)}{d\chi}$ and $\tilde{\beta}(\chi) = \frac{d^2\textup{ln}A(\chi)}{d\chi^2}$. These equations represent a system of coupled, second order wave equations for the perturbations $\zeta$ and $\delta\chi$ and in the $H(\chi)=\sin(\chi)$ case they reduce to the ones in \cite{Doneva:2020stability_tns}. The perturbations of the metric functions, the energy density and the pressure in terms of $\zeta$ and $\delta\chi$ are as follows:
	
	\begin{widetext}
		\begin{align}
			&\delta\Lambda=a^2r\chi_0'\delta\chi - 4\pi G_*A^4(\chi_0)(\tilde{\varepsilon}_0+\tilde{p}_0)e^{2\Lambda_0}r\zeta, \\
			&\delta\tilde{\varepsilon}=-(\tilde{\varepsilon}_0+\tilde{p}_0)\left[r^{-2}e^{-\Lambda_0}\left(e^{\Lambda_0}r^2\zeta\right)'+\delta\Lambda\right] - [\tilde{\varepsilon}_0'+3\alpha(\chi_0)(\tilde{\varepsilon}_0+\tilde{p}_0)\chi_0']\zeta - 3\alpha(\chi_0)(\tilde{\varepsilon}_0+\tilde{p}_0)\delta\chi, \\
			&\delta\tilde{p}=\tilde{c}^2_s\delta\tilde{\varepsilon}, \\
			&\delta\Gamma'=\frac{1}{r}\left[1 - 2 a^2 H(\chi)^2+r^2\left(8\pi G_*A^4(\chi_0)\tilde{p}_0-2V(\chi_0)\right)\right]e^{2\Lambda_0}\delta\Lambda \notag \\
			&\qquad + re^{2\Lambda_0}\left[-\partial_\chi V(\chi_0)-2\frac{a^2}{r^2}H(\chi)\frac{d}{d\chi}H+ 16\pi G_*\alpha(\chi_0)A^4(\chi_0)\tilde{p}_0\right]\delta\chi + a^2r\chi_0'\delta\chi'+4\pi G_*e^{2\Lambda_0}r A^4(\chi_0)\delta\tilde{p},
		\end{align}
	\end{widetext}
	where $\tilde{c}^2_s$ is the sound speed in the Jordan frame and is defined by $\tilde{c}^2_s=\frac{d\tilde{p}_0}{d\tilde{\varepsilon}_0}$. 
	
	The boundary conditions  at the center of the star are derived from the requirement for regularity of the perturbations and we have $\zeta(t,r=0)=0$ and $\delta\chi(t,r=0)=0$. Similar to pure GR case, the Lagrangian perturbation of the pressure $\Delta\tilde{p}$ has to vanish at the surface of the star. Only the perturbation of the scalar field $\delta\chi$ can propagate outside the star while $\zeta$ vanishes there. For large distances $\delta\chi$ has to satisfy the radiative (outgoing) asymptotic condition, expressed as
	\begin{equation} \label{radiative asymptotic condition}
		\partial_t(r\delta\chi)+\partial_r(r\delta\chi)=0.
	\end{equation}
	
	\subsection{Results linear stability}
		\begin{figure}
		\centering
		\includegraphics[scale=0.38]{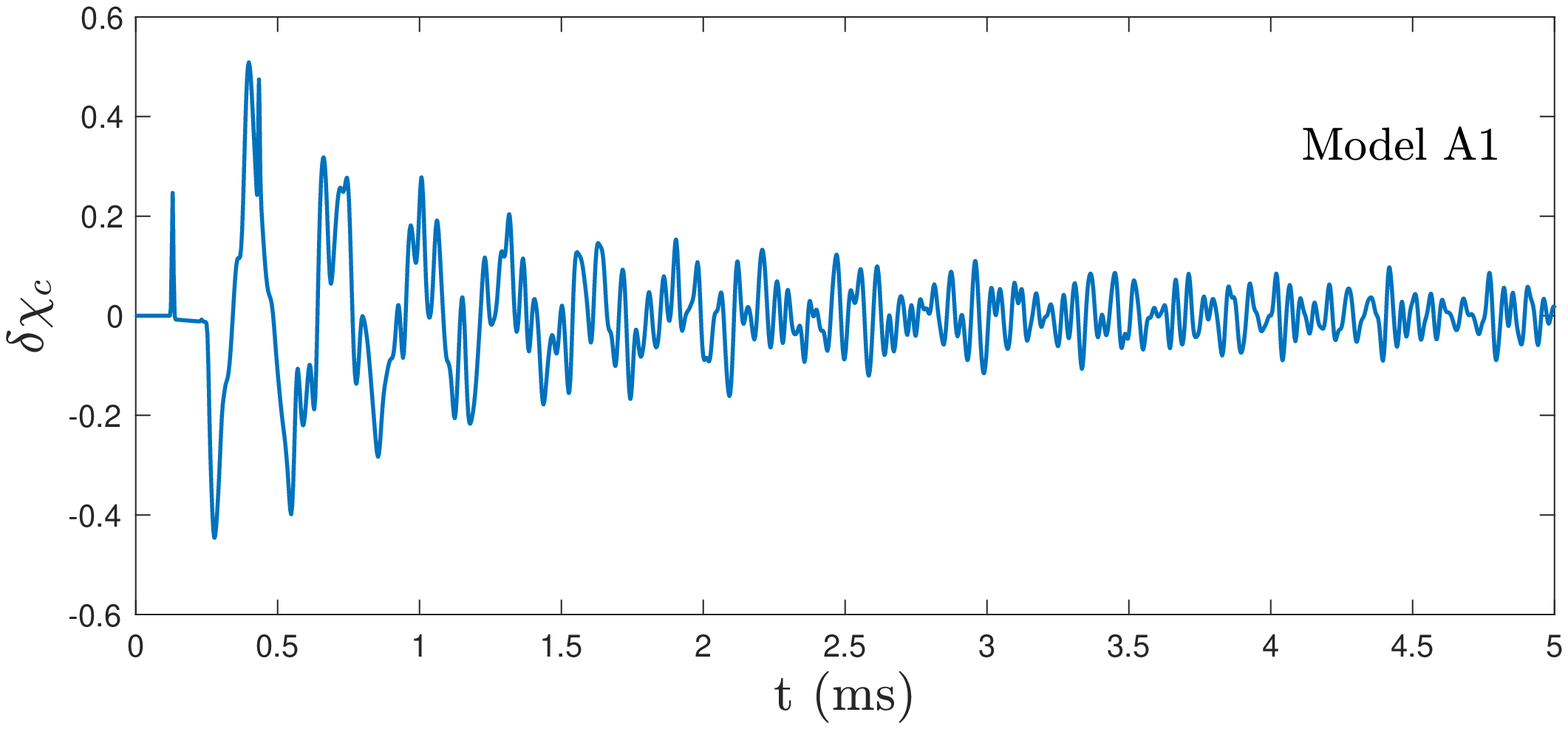}
		\includegraphics[scale=0.38]{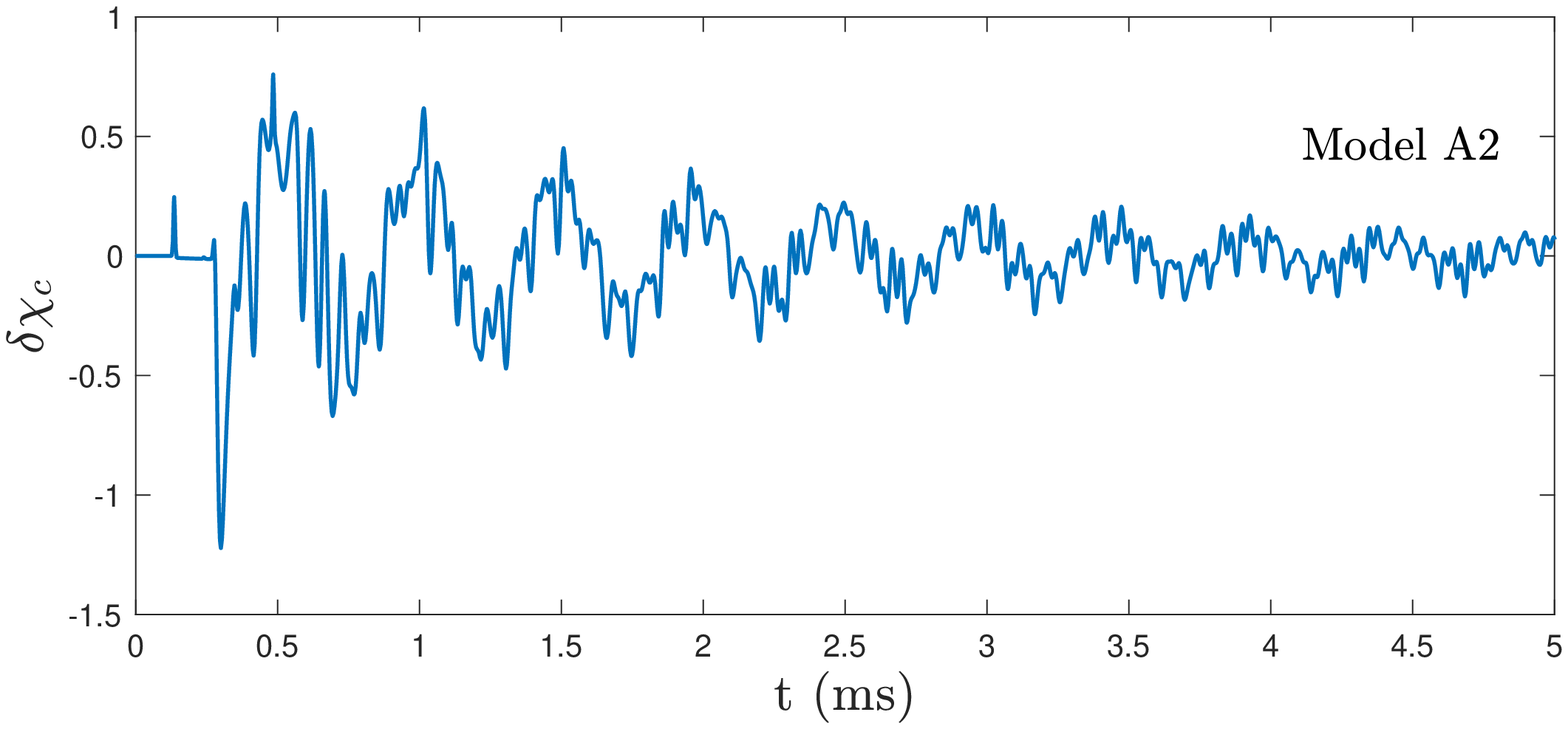}
		\includegraphics[scale=0.38]{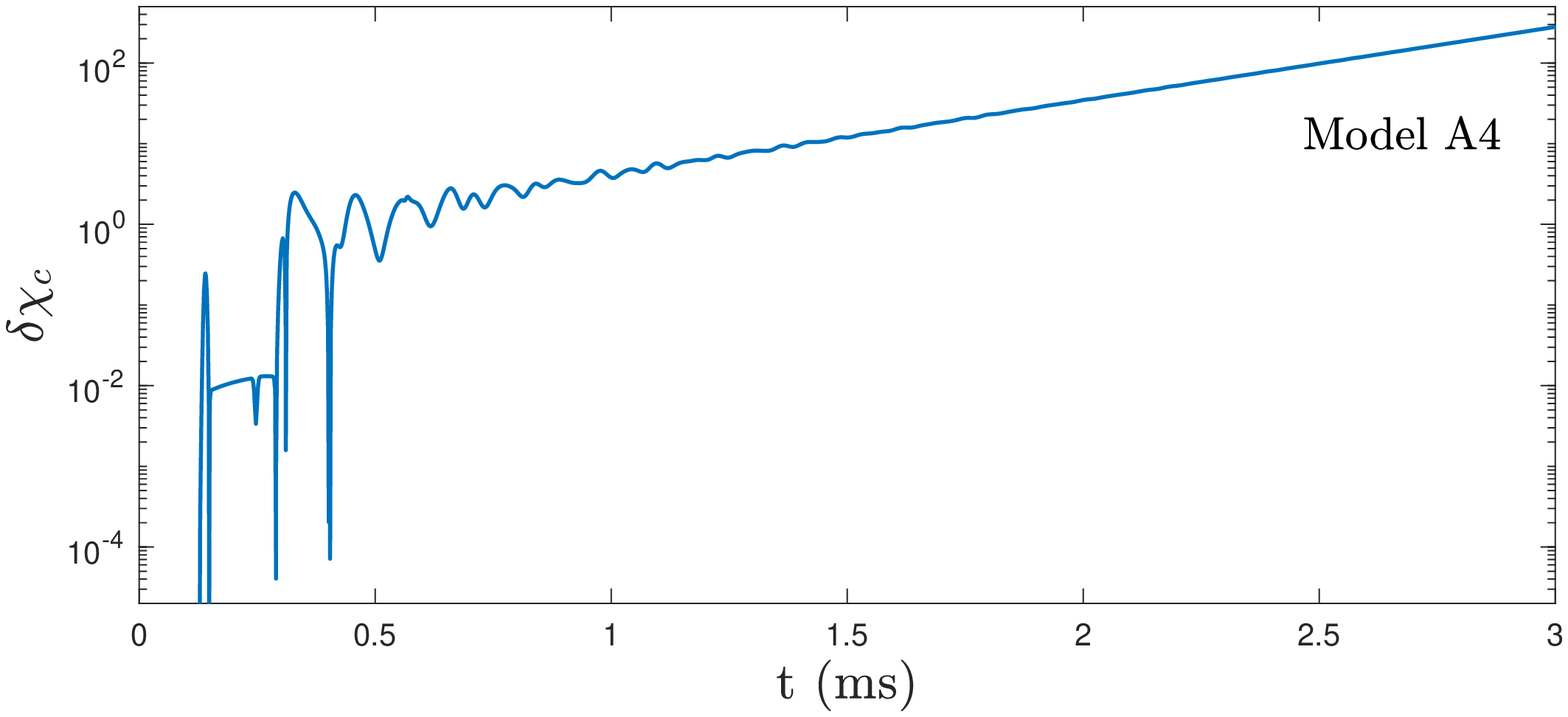}
		\caption{The evolution of the scalar field perturbations $\delta\chi_{c}$ extracted at a point located several neutron star radii away from the surface, with $A(\chi)=e^{\beta\sin^2{\chi}}$, $H(\chi)=\sin{\chi}$, $a^2=0.1$ and $\beta=-1.5$. Three models are considered, including two stable (top and middle panel) and one unstable (bottom panel). These are models A1, A2 and A4 listed in Table \ref{tab:NSdata} and discussed in detail in the next section.
		}
		\label{fig:dchi_waveforms}
	\end{figure}

	To perform the stability analysis in the linear regime, we convert the linearized wave equations \eqref{eq:WE1} and \eqref{eq:WE2} into a form more suitable for numerical analysis by adapting a standard approach from pure GR \cite{Doneva:2020stability_tns, Ruoff:2000}. Namely, we introduce a new dimensionless function
	\begin{equation}
		Z(t,r)=(\Tilde{\varepsilon}_0+\Tilde{p}_0)r\zeta e^{2\Lambda_0}.
	\end{equation}
	Since this function is zero at the stellar surface where $\Tilde{\varepsilon}_0$ and $\Tilde{p}_0$ vanish, applying the boundary conditions is easier in terms of $Z$. 
	
	To evolve the perturbation equations \eqref{eq:WE1} and \eqref{eq:WE2}  in time, we use the Leapfrog method. As initial data for $\delta\chi$ we use a Gaussian pulse which is located several neutron star radii away from the stellar surface, with zero initial velocity at $t=0$. $Z$ is set to be zero initially and is always zero outside the star by construction. It will remain zero until the $\delta\chi$ pulse reaches the star and will get excited only then through the coupling of the fluid and scalar field perturbations. 
	
	Using the method described above, we solved the system of equations for different forms of $H(\chi)$ and $A(\chi)$ for different values of $\beta$. We found that for all of the considered scalarized neutron star branches, the perturbation $\delta\chi$ decays in time for the scalarized models before the maximum of the mass, which implies the branch is stable up to this point \footnote{Let us point out that contrary to the GR case, the radial oscillations in TMST will have an amplitude decaying in time because the scalar field carries away energy to infinity.}. Whereas, for neutron star models located after the maximum of the mass $\delta\chi$ grows exponentially which clearly indicates instability.  For smaller values of $\beta$ (for example the $\beta=-1.5$ branch in Fig. \ref{fig:M_rhoc}) an interesting observation has been made. In the region where two scalarized solutions exist for the same central energy densities, it was found that both solutions are stable. This implies that at these central energy densities, three radially stable solutions exist simultaneously: one general relativistic and two scalarized solutions. Even more interestingly, in the part of the scalarized branch just after the bifurcation point, the mass increases with decreasing central energy density while  the neutron stars is still stable.
	
	Fig. \ref{fig:dchi_waveforms} shows the waveforms of the central scalar perturbation $\delta\chi_{c}$  for three representative scalarized neutrons stars  belonging to the $\beta=-1.5$ branch in Fig. \ref{fig:M_rhoc}. The \emph{top} figure depicts $\delta\chi_{c}$ of a star from the initial part of the branch where the mass increases with decrease of the central energy density. The \emph{middle} figure refers to a star from the part of the branch where the mass increases as central energy density increases, but having an energy density smaller than the solution with maximum mass. Finally, the \emph{bottom} figure represents the perturbations of an unstable star with central energy density slightly higher than the maximum mass solution. As one can see, after a few milliseconds the perturbation function $\delta \chi$ shooting off exponentially. The time at which instability sets in reduces for stars with higher central energy density. Here we will not comment in detail on the frequencies of the radial oscillations since the focus of the paper is on the stability, but our analysis shows that, as expected, these frequencies decrease monotonically with the increase of the stellar mass and they cross zero exactly for the maximum mass models.

	\section{Nonlinear Scheme}
	\label{section:nonlinear}
	
	Having done the linear analysis of the stability of scalarized models, we now turn to address the issue within fully non-linear framework. Among the advantages of the non-linear analysis is that one can access more information about how the instabilities grow and saturate. As a whole, the evolutionary equations in TMST (Sec.~\ref{sec.V.A}) resemble those in DEF theories with some additional terms owing to the non-trivial geometry of target spaces. It thus justifies the appliance of the numerical approach (reconstruction method and high-performance-shock-capture algorithm) that has been implemented in DEF theories in  \cite{OConnor:2009iuz,Gerosa:2016fri} to TMST. We construct a grid adequate for our purpose in this work (Sec.~\ref{sec.V.B}) for solving the evolutionary equations. It has been checked, that the results summarized in Sec.~\ref{sec.V.C} show only slight deviations by doubling the resolution.
	
	\subsection{Evolution Equations}\label{sec.V.A}
	
	The Euler equation,
	\begin{align}\label{eq:euler}
		\nabla_{\mu}\tilde{T}^{\mu\nu}=0,
	\end{align} 
	can be presented as a first-order flux conservative system \cite{Banyuls:1997zz,Font:2000pp}, 
	\begin{align}\label{eq:fluxcon}
		\partial_t \textbf{U} + \frac{1}{r^{2}}\partial_{r} \left[ r^{2}
		\frac{\alpha}{X} \textbf{f(U)} \right] = \textbf{s(U)},
	\end{align}
	constituting the \emph{conserved} quantities $\boldsymbol{U}=\{ D,\tau,S^{r)}\}$ and the corresponding fluxes \textbf{f(U)} and sources \textbf{s(U)}.
	The Jacobian of this (differential equation) system, $\frac{\partial {\bf f(U)}}{\partial {\bf U}}$, offers information about the characteristic speeds of the \emph{conserved} quantities.
	Defining the \emph{conserved} quantities and the fluxes via
	\begin{subequations}
		\begin{align}
			D &= \frac{A^4e^{\Lambda}  }{\sqrt{1-v^2}} \tilde{\rho},  \\
			S^r &= \frac{ A^4 v}{1-v^2}  (\tilde{\varepsilon}+\tilde{p}), \\
			\tau  &= \frac{A^4\tilde{\varepsilon}}{(1-v^2)} - A^4\tilde{p}-D,
		\end{align}
	\end{subequations}
	and 
	\begin{subequations}
		\begin{align}
			f_{D} &= {D} v, \\
			f_{S^{r}} &= {S}^r v + A^4\tilde{p}, \\
			f_{\tau} &= {S}^r-{D} v,
		\end{align}
	\end{subequations}
	we find the source terms
	\begin{widetext}
		\begin{subequations}
			\begin{align}
				s_{D} =&D e^{\Gamma} ( \psi+\eta v) A\dF, \\
				s_{S^{r}} =& ({S}^r v -{\tau} -{D})  e^{\Gamma+\Lambda} \left(8\pi r A^4\tilde{p} + \frac{m}{r^2}
				+ e^{-\Lambda} A\dF\eta   -rV_{\text{eff}}	  \right)   + e^{\Gamma+\Lambda}  \frac{A^4\tilde{p}m}{r^2}
				+ 2e^{\Gamma-\Lambda}\frac{A^4\tilde{p}}{r} - 2re^{\Gamma+\Lambda} {S}^r \eta \psi A^2 a^{2} \nonumber\\
				&+ 3e^{\Gamma} A^5\tilde{p} \dF \eta 
				-e^{\Gamma+\Lambda} A^4\tilde{p}rV_{\text{eff}}   - \frac{r}{2}e^{\Gamma+\Lambda}(\eta^2+\psi^2) \left( {\tau} + A^4\tilde{p}  + {D} \right) (1+v^2)A^2 a^{2}, \\
				s_{\tau} =& -\left( {\tau}+ A^4\tilde{p} + {D} \right) re^{\Gamma+\Lambda}
				\bigg( (1+v^2) \eta \psi + v (\eta^2 + \psi^2) \bigg) A^2a^{2}   -e^{\Gamma}A\dF \left[ {D} v \eta + \left( {S}^rv - {\tau}+ 3A^4\tilde{p}  \right) \psi \right].
			\end{align}
		\end{subequations}
	\end{widetext}
	In addition, it has been illustrated in \cite{Gerosa:2016fri} that the characteristic speeds, determined by {\bf f(U)} and {\bf U}, for the conservative system in DEF theories are exactly the same as those in GR due to their independence on the coupling function $A$. In our formulation for TMST, we stick with the same definition of {\bf f(U)} and {\bf U} as \cite{Gerosa:2016fri}, indicating that the characteristic properties for the system \eqref{eq:fluxcon} are identical to GR.
	
	Having assumed $\Theta=\theta$ and $\Phi=\phi$, the nonlinear evolution equation for the scalar fields reads
	\begin{align}\label{eq:scalwv}
		\square\chi-\frac{2H}{r^{2}}\frac{\partial H}{\partial\chi}-\frac{1}{a^{2}}\frac{\partial V}{\partial\chi}
		=-\frac{4\pi}{a^{2}}\frac{\partial \ln A}{\partial\chi} T,
	\end{align}
	which can be reduced to two first order, decoupled equations having the form
	\begin{subequations}\label{eq:scalareom}
		\begin{align}
			\dot{\eta}=&\frac{e^{-\Lambda}}{A}(Ae^{\Gamma}\psi)'
			-re^{\Gamma+\Lambda}\eta\bigg(a^{2}A^2 \psi\eta-4\pi s^{r}\bigg) 
			- \psi\eta e^{\Gamma} A\dF, \\
			\dot{\psi}=&\frac{e^{-\Lambda}}{Ar^{2}} (Ae^{\Gamma} r^{2}\eta)'
			- re^{\Gamma+\Lambda}\psi\bigg( a^{2}A^2\psi\eta-4\pi s^{r} \bigg)
			- \psi^{2}e^{\Gamma}A\dF \nonumber\\
			& - \frac{4\pi e^{\Gamma}}{Aa^{2}}\dF
			\bigg( \tau - s^{r}v+D-3A^4 \tilde{p}  \bigg)
			-  \frac{e^{\Gamma}}{Ar^{2}a^{2}}  \frac{d}{d\chi} \bigg( r^{2} V_{\text{eff}}  \bigg),
		\end{align}
	\end{subequations}
	with $\psi=e^{-\Gamma}\dot{\chi}$ and $\eta=e^{-\Lambda}\chi'$. 
	The effective potential is defined as
	\begin{align}
		V_{\text{eff}}=V+\frac{a^{2}H^{2}(\chi)}{r^{2}},
	\end{align}
	where the second term on the right hand side attributes to the geometry of the target space manifold.
	The Einstein equations reduce to two linearly independent equations,
	\begin{subequations}
		\begin{align}
			\Gamma' &= e^{2\Lambda}\bigg[ \frac{m}{r^{2}}+4\pi r\bigg(s^{r}v+A^{4}\tilde{p}\bigg)
			+\frac{a^{2}r}{2}A^2(\psi^{2}+\eta^{2})
			-rV_{\text{eff}} \bigg], \\
			m' &= 4\pi r^{2}(\tau+D)+\frac{  a^{2}r^{2} } {2} A^2(\psi^{2}+\eta^{2})
			+r^{2} V_{\text{eff}},
		\end{align}
	\end{subequations}
	relating the spatial derivative of the metric functions to the fluid quantities and the scalar field.

	\subsection{Numerical setup}\label{sec.V.B}
	
	The code used in this work to solve the above system of nonlinear evolution equations is a modification of the \texttt{GR1D} code \cite{OConnor:2009iuz,OConnor:2014sgn} (for the DEF theory version of \texttt{GR1D}, readers can refer to, e.g., \cite{Gerosa:2016fri,Sperhake:2017itk,Cheong:2018gzn,Rosca-Mead:2019seq}). In this spherical symmetric simulation, the computational domain ranges from the stellar center to $r=10000$ km ($\sim 1000$ times the radius of the star), securing that the radial boundary is sufficiently far away from the strong-field region where the spacetime is well approximated by Minkowski metirc. The grid used has uniform size of $30$ m from center to $r=40$ km and the grid size increases exponentially from $r=40$ km toward the outer boundary in the rate that the number of grid points amounts to 10000. There are, therefore, $\sim 330$ grids point  inside stars. 
	At the center and the outer boundary, the boundary conditions are applied to every metric functions and fluid variables. The radial velocity $v$ is antisymmetric across the origin since the radial fluxes vanish there, while the remaining variables are symmetric. All variables are symmetric about the outer edge. 
	
	We do not perturb artificially any quantities ($\Gamma$, $\Lambda$, $\chi$, ...), but only the error due to numerical truncation serves as perturbation to the equilibrium.

	\subsection{Results} \label{sec.V.C}
	
	We examine the stability of scalarized neutron stars along the sequences of equilibrium models depicted in Fig.~\ref{fig:M_rhoc}. To balance the completeness of our results and the compactness of this paper, we choose without loss of generality some symbolic models with $H=\sin\chi$ to illustrate our results, whose properties are listed in Table.~\ref{tab:NSdata}. 
	
\begin{figure}
		\centering
		\includegraphics[scale=0.45]{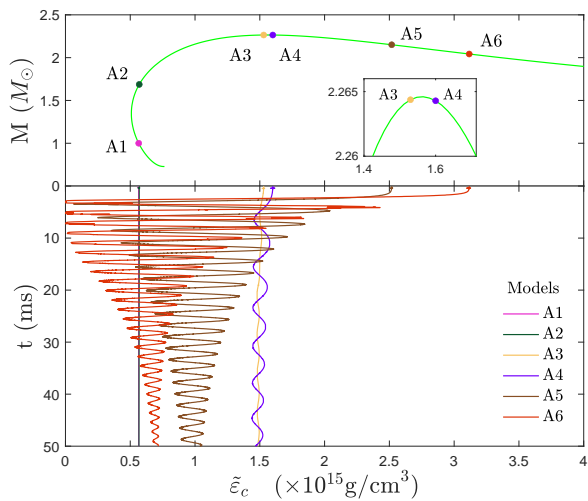}
		\caption{\emph{top}: The non-GR part of the sequence of solutions with $H(\chi)=\sin\chi$, $\beta=-1.5$ and $a^{2}=0.1$, along which the six models A1-6 in Tab.~\ref{tab:NSdata} are marked by different colors. The magnified window shows that A3 and A4 models have, respectively, slightly smaller and larger mass than the maximal mass, where the instability kinks in.
			\emph{bottom}: Evolutions of central energy density of A1-6. 
		}
		\label{fig:evolutionchartA}
	\end{figure}

	In Fig.~\ref{fig:evolutionchartA}, we summarize the evolution of $\tilde{\varepsilon}_{c}$ of models A1-A6, where each history is arranged in the order of the initial values of $\tilde{\varepsilon}_{c}$. 
	The models A1 and A2 oscillate about the equilibrium slightly, whereas the model A3 shifts a bit toward left and oscillates around a non-zero residual with respect to its initial value, which converges to zero as second order with increasing resolution.
	The results for A1-3 reflect that the segment, which is non-GR and yet reaches the maximal mass, is stable.
	The stability is lost when the maximal mass is reached; particularly, model A4 exhibits instability and deforms into a stable model. The point representing A4 on upper panel of Fig.~\ref{fig:evolutionchartA} drifts toward left then oscillates around another point on the curve with the same baryon mass, as expected. The unstable models A5 and A6 also show the deformation into a stable model with same baryon mass. 
	
\begin{figure}
		\centering
		\includegraphics[scale=0.43]{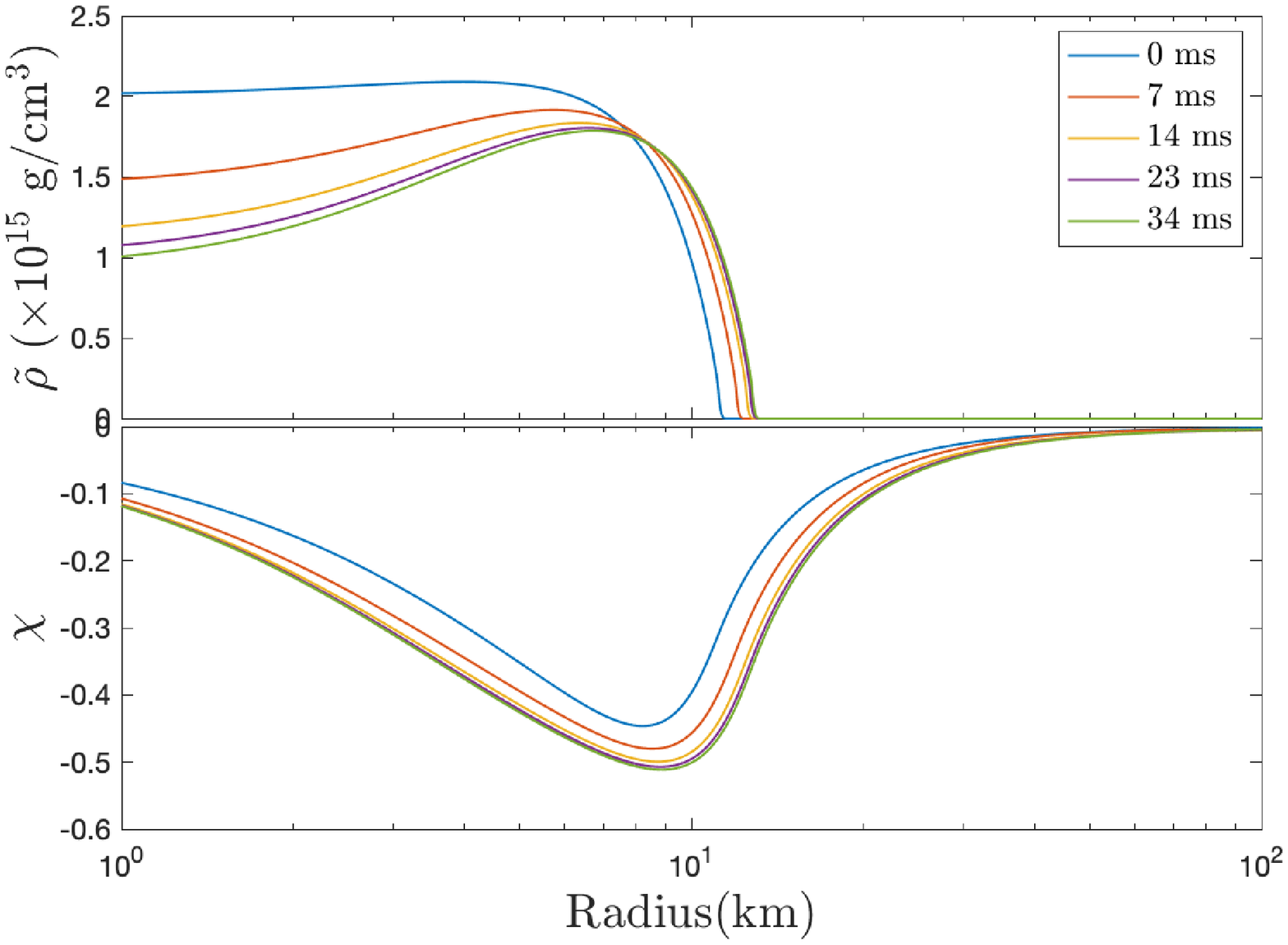}
		\includegraphics[scale=0.45]{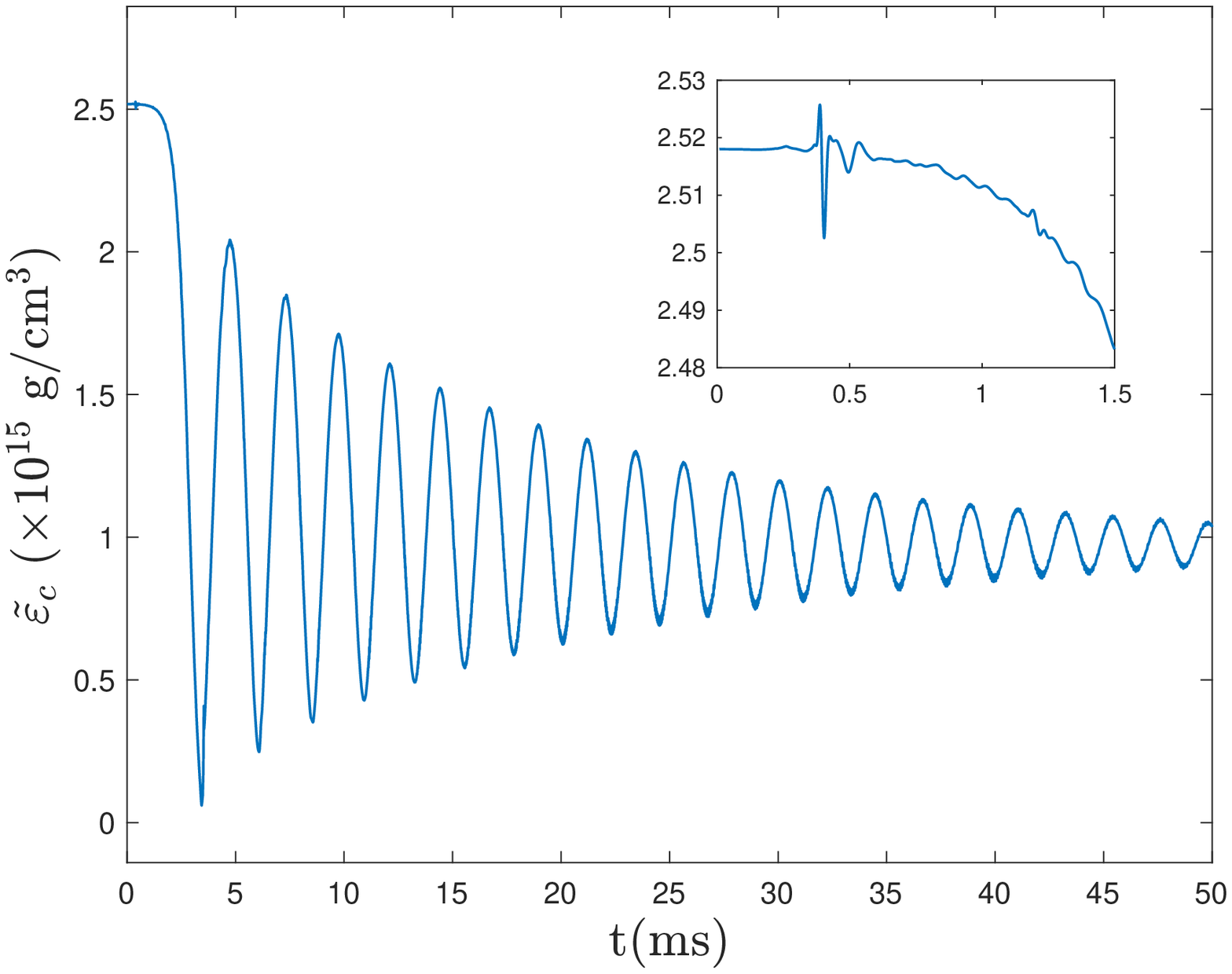}
		\caption{ \emph{top}: Distributions of the baryon mass density $\tilde{\rho}$ and the scalar field $\chi$ at several moments, whereby one can see that the material part of star settles to the final state at $\sim 34$ ms, while $\chi$ has already reached to the final profile at $\sim 23$ ms.
		\emph{bottom}: For model A5 the central energy density ${\tilde \varepsilon}_c$ is plotted as a function of time. In the magnified window, the onset of instability is shown.
		Model A5 has been considered for both panels.
		}
		\label{fig:drift}
	\end{figure}

\begin{figure}
		\centering
		\includegraphics[scale=0.45]{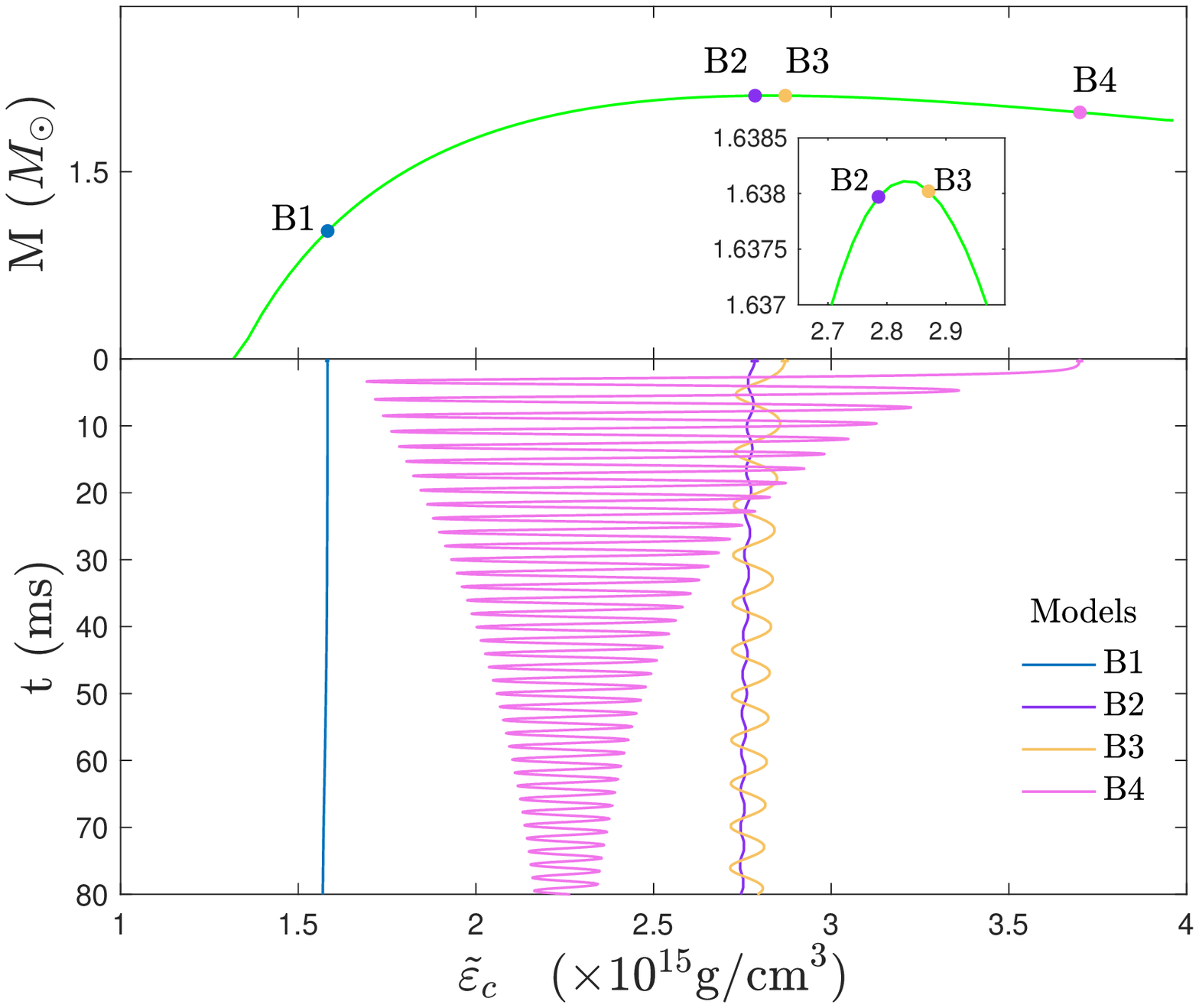}
		\caption{\emph{top}: Stationary solutions with $H(\chi)=\sin\chi$, $\beta=-1$ and $a^{2}=0.1$. 
			The marks represent the four models B1-4 in Tab.~\ref{tab:NSdata}. The magnified window shows that B2 and B3 models have, respectively, slightly smaller and larger mass than the maximal mass, where the instability kinks in.
			\emph{bottom}: Evolution for models B1-4. 
		}
		\label{fig:evolutionchartB}
	\end{figure}

	In particular, the translation of model A5 from the initial unstable star to the stable one is shown by the evolution of the radial profiles of the baryon density $\tilde{\rho}$ and the scalar field $\chi$ (top panel of Fig.~\ref{fig:drift}). One can observe that the material part of the star settles to the final state at $\sim 34$ ms, while $\chi$ has already reached to the final profile at $\sim 23$ ms. The development of the instability is depicted by the evolution of $\tilde{\varepsilon_{c}}$ and the central value of scalar field $\chi_{c}$ (bottom panel of Fig.~\ref{fig:drift}), where the magnified windows show the onset of the instability and the following saturation is apparent in the main figure.
	On the other hand, the evolution of models B1-B4 are given in Fig.~\ref{fig:evolutionchartB}, which confirms as well that the non-GR segment left to the maximal mass, is stable and the segment right to the maximal mass is unstable.
	We note that in general an unstable neutron star could migrate to a stable star with same baryon mass but less compact, or collapse to a black hole, i.e. there should be a third channel that an unstable star collapses into a BH.
	However, our tests show that this channel probably requires additional perturbation.

	\begin{table}
		\caption{Properties of symbolic models with the target space geometry $H(\chi)=\sin\chi$ and the coupling function $A(\chi)=\exp(\beta\sin^{2}\chi/2)$.
			There are two classes of the chosen models, where models in the ``A'' class are solutions for $\beta=-1.5$ and models in the ``B'' class are solutions for $\beta=-1$.
			The second toward the final columns are, respectively, the central energy density, the radius, and the (baryon) mass of stars. }
		\begin{tabular}{c|c|c|c}
			\hline
			\hline
			Model & $\tilde{\varepsilon}_{c}$ (g/cm$^{3}$) & Radius (km) & Mass ($M_{\odot}$)  \\
			\hline
			A1 & 5.65364$\times 10^{14}$ & 10.0876 & 1.00017 \\
			A2 & 5.69327$\times 10^{14}$  & 10.9379 & 1.68717 \\
			A3 & 1.52979$\times 10^{15}$ & 11.0621 & 2.04260 \\
			A4 & 1.60004$\times 10^{15}$ & 11.0094 & 2.26437 \\
			A5 & 2.51809$\times 10^{15}$ & 10.2366 & 2.26429 \\
			A6 & 3.11633$\times 10^{15}$  & 9.7595 & 2.15068 \\
			\hline
			B1 & 1.58276$\times 10^{15}$  & 10.4074 & 1.39231 \\
			B2 & 2.78566$\times 10^{15}$ & 9.5503 & 1.63797 \\
			B3 & 2.87078$\times 10^{15}$ & 9.4966 & 1.63802 \\
			B4 & 3.69972$\times 10^{15}$ & 9.0108 & 1.60768 \\
			\hline
			\hline
		\end{tabular}
		\label{tab:NSdata}
	\end{table}
	
	Since the solutions to $H(\chi)=\sin\chi,\chi$, and $\sinh\chi$ differ only quantitatively while remaining qualitatively the same, it is expected that the stability properties for each branches do not change among these three choices of $H(\chi)$. 
	In practice, we confirm this hypothesis by analyzing the stability of some representative models of each branches, and conclude the same -- a model lighter than the maximal mass is stable, otherwise is unstable.
	
	Having evolved and checked stability for a large number of models, we find that each sequence contains exactly one stable segment and one unstable, converging towards the maximum mass models. The non-GR parts of the stable segments for $\beta=-1.5$ can be further divided into two classes: one before the central energy density reaches the minimal value, and one after. It is of particular interest that the scalarized models belonging to the part of the branch before the minimal $\tilde{\varepsilon}_{c}$ are stable even though they have larger masses for smaller $\tilde{\varepsilon}_{c}$. It indicates roughly that these models are ``glued together'' more by the non-trivial scalar field rather than by the self-gravitating fluid. In some sense, this is also the reason why the maximal mass of the solutions in TSMT (also in DEF theories) is larger than the predicted one by GR.

	\section{Conclusions}
	\label{section:concl}
	
	In the present paper we have investigated the (in)stability of scalarized neutron stars in tensor-multi-scalar theories. These models posses two very intriguing properties. First, their scalar charge is vanishing leading to zero scalar dipole radiation. Therefore, no constraints can be imposed by the binary pulsar observations, contrary to the DEF model in standard scalar tensor theories. Second, there exists a region of non-uniqueness of the scalarized solutions themselves, i.e. for a certain range of central energy densities two scalarized solutions can co-exist with the GR (zero scalar field) one. Clearly, this interesting structure calls for an investigation of the stability. We used two approaches in order to be able to confirm independently the results -- solving the linearized perturbation equation and addressing the full nonlinear evolution in spherical symmetry. The equations governing the evolution of the scalar field and the metric were derived independently in the considered class of tensor-multi-scalar theories and they were solved numerically.
	
	The linear stability analysis showed, that for all combinations of parameters we have studied, the critical point for stability  occurs at the maximum of the mass. Thus the scalarized branches before this point are stable, independent on whether they posses a region of non-uniqueness in terms of the central energy density or not. This is a very interesting conclusion leading to the fact that there is a part of the branch where the total mass of the neutron stars increases with the decrease of the central energy density that is in sharp difference with GR and even with most of the known alternative theories of gravity.  As expected, the GR solutions with trivial scalar field loose stability at the point of the first bifurcation. Their stability is restored once the scalarized branch merges again with the GR one (only in case the second bifurcation point is before the maximum mass of the GR sequence of course). 
	
	 In the fully non-linear investigation, we again identified the parts on the sequence of scalarized models that are unstable and the results agree perfectly with the ones from the linear perturbation analysis. 
	 A particular merit of the non-linear treatment is that apart from demonstrating the development of the instability we can follow the evolution towards a final stable state. The transition from an unstable model to a stable one with the same baryon mass is numerically revealed in our simulations. However, the dynamics (damping timescale of instabilities, the emission via the scalar channel during the drift from an unstable model to a stable one, etc.) behind the phenomenon is not addressed in the present work. The knowledge of the detailed dynamics is crucial in connecting the instabilities of the objects discussed here to observations, thus research towards this direction will be helpful providing possible constraints on TMST.

	\section*{Acknowledgements}
	HJK appreciates the financial support of the Sandwich grant (JYP) No.~109-2927-I-007-503 by DAAD and MOST. JS would like to acknowledge the support from Trieste-Chalmers Ph.D. Fellowship and the computing centre of INAF-Osservatorio Astronomico di Trieste, under the coordination of the CHIPP project \cite{bertocco2019inaf, taffoni2020chipp}.
	DD acknowledges financial support via an Emmy Noether Research Group funded by the German Research Foundation (DFG) under grant
	no. DO 1771/1-1.  SY would like to thank the University of T\"ubingen for the financial support. The partial support by the Bulgarian NSF Grant DCOST 01/6 and the  Networking support by the COST Actions  CA16104 and CA16214 are also gratefully acknowledged.
	

\end{document}